\documentclass[aps,reprint,superscriptaddress]{revtex4-1}
 \usepackage[utf8]{inputenc} 

\newcommand{\bea}{\begin{eqnarray}}
\newcommand{\eea}{\end{eqnarray}}
\usepackage{tikz}
\usepackage{upgreek}
\usepackage{bbold}
\usepackage{amsmath}

\begin{document}

\title{Euclidean operator growth and quantum chaos}

\author{Alexander Avdoshkin}
\affiliation{Department of Physics, University of California, \\
Berkeley, CA 94720, USA
}
\author{Anatoly Dymarsky}
\affiliation{Skolkovo Institute of Science and Technology,\\
Moscow, Russia, 143026}
\affiliation{Department of Physics and Astronomy, \\
University of Kentucky, Lexington, KY, 40506}

\date{\today}

\begin{abstract}
We consider  growth of local operators under Euclidean time evolution in lattice systems with local interactions. 
We derive rigorous bounds on the operator norm growth and then proceed to establish an analog of the Lieb-Robinson bound for the spatial growth. In contrast to the Minkowski case when ballistic spreading of operators is universal, in the Euclidean case spatial growth is system-dependent and indicates if the system is integrable or chaotic. In the integrable case, the Euclidean spatial growth is at most polynomial. In the chaotic case, it is the fastest possible: exponential in 1D, while in higher dimensions and on Bethe lattices local operators can reach spatial  infinity in finite Euclidean time. We use bounds on the Euclidean growth to establish constraints on individual matrix elements and operator power spectrum. We show that one-dimensional systems  are special with the power spectrum always being superexponentially suppressed at large frequencies. Finally, we relate the bound on the Euclidean growth to the bound on  the growth of Lanczos coefficients. To that end, we develop a path integral formalism for the  weighted Dyck paths and evaluate it using saddle point approximation. Using a conjectural connection between the growth of the  Lanczos coefficients and the Lyapunov exponent controlling the growth of OTOCs, we propose an improved bound on chaos valid at all temperatures. 
 
\end{abstract}

\maketitle

\section{Introduction and Results}
Operator spreading, or growth, in local systems is a question of primary interest, which encodes transport properties, emergence of  chaos  and other aspects of many-body quantum dynamics \cite{Roberts:2014isa,Roberts:2014ifa,Nahum:2017yvy,Khemani:2017nda,Qi:2018bje,Bentsen_2019,Parker_2019,Barbon:2019wsy}. A classic result of Lieb and Robinson \cite{LR} (see also  \cite{Huang_2018,chen2019operator} for recent progress) establishes that under  time evolution the  fastest possible  spatial spreading of local operators is  ballistic. There is no norm growth in this case since the time evolution is unitary. Ballistic spreading of operators, and signals, has been established for many models \cite{Calabrese:2006rx,ALEINER2016378,Luitz_2017,Nahum:2017yvy,Patel2017vfp,das2018light, Tibor1, Tibor2} and seems to be a universal feature of  local systems in any dimensions.  At the same time, evolution of local operators in Euclidean time 
\bea
A(-i\beta)=e^{\beta H}A\, e^{-\beta H}, \label{Agr}
\eea
which we study in this paper, is much more nuanced. Since the Euclidean evolution is not unitary, the norm of $A(-i\beta)$ quickly grows with $\beta$. Moreover, as we explain below, the operator growth is not universal and reflects if the system in question is integrable or chaotic.

We start in section \ref{sec:normbound} by deriving a bound on $|A(t)|$  valid uniformly for $|t|=\beta$ by expanding \eqref{Agr} in Taylor series and bounding corresponding nested commutators. For local $H$ there is a combinatorial problem of counting contributing nested commutators, which we solve exactly for short range systems defined on Bethe lattices, which includes local systems in 1D.   In higher dimensions we conjecture an asymptotically  tight bound. Hence, we expect  our bounds on operator norm to be optimal in the class of Hamiltonians we consider  --  lattice Hamiltonians with local interactions. We find that maximal rate of growth is very different in 1D, where it  is at most double-exponential, and in higher dimensions or Bethe lattices, where the norm can become infinite in finite Euclidean time. 
We extend the analysis to include spatial growth in section \ref{sec:LB}, where we find that in 1D  operators spread at most exponentially, while in higher dimensions, including Bethe lattices, they can reach spatial infinity in finite Euclidean time. When the 1D system is finite, the minimal time necessary for an operator to reach the boundary is logarithmic, which may explain logarithmic convergence  of the numerical Euclidean time algorithm proposed in \cite{Beach_2019}. We further speculate in section \ref{sec:S} that the timescale originating from the Euclidean  Lieb-Robinson bound might be related to the Thouless energy of the corresponding quantum many-body  system \cite{Chan_2018}.

In section \ref{sec:ME}  the results on norm growth are used to constrain individual matrix elements. We find that matrix elements in energy eigenbasis $\langle E_i|A|E_j\rangle$ must decay at least exponentially with $\omega=|E_i-E_j|$, while in 1D the decay must be faster than exponential, as provided by \eqref{meb} and \eqref{bme}. We also establish a number of bounds on the  auto-correlation function at finite temperature $C_T(t)$, and its Fourier transform -- the power spectrum $\Phi_T(\omega)$, 
\bea
\label{Cdef}
C_T(t)\equiv {\rm Tr}(\rho\, A(t) A)=\int\limits_{-\infty}^\infty
\Phi_T(\omega)e^{i\omega t}d\omega, \\ \nonumber
\rho \propto e^{-H/T},\, {\rm Tr}(\rho)=1. 
\eea
The bounds have integral form, see (\ref{mu1d},\ref{mubound}) and \eqref{abaninbound}. At the physical level of rigor, they suggest that $\Phi_T(\omega)$ decreases exponentially with $\omega$ in $D\geq 2$, while in 1D the decay at large frequencies is superexponential. This emphasizes that one-dimensional systems are indeed very special, and many numerical results established for one dimensional systems may not necessarily apply to higher-dimensional systems.  

The bound on $|A(t)|$ established in section \ref{sec:normbound} depends only on the absolute value $|t|$. Obviously, it is overly conservative for real $t$ when the time evolution is unitary. We argue, however, in section \ref{sec:S} that it does correctly capture the Euclidean growth $t=-i\beta$ of chaotic systems. We also consider system size dependence of $|A(t)|$ and find it to be consistent with the Eigenstate Thermalization Hypothesis (ETH). 
For the integrable systems we find the growth of $|A(t)|$ to be much slower than maximal possible, and in particular spatial growth of $A(-i\beta)$ in this case is not  exponential but polynomial. 

The bound on $|A(-i\beta)|$ can be translated into a bound on the growth of  Lanczos coefficients $b_n$, appearing as a part of the recursion method to numerically compute $C_T(t)$. This is provided we {\it assume} that asymptotically $b_n$ is a smooth function of $n$. To perform this calculation, we introduce a formalism of summing over weighted Dyck paths in section \ref{sec:L}, and evaluate the corresponding path integral via saddle point approximation. 

The obtained bound on Lanczos coefficients  growth \eqref{boundonLT} is valid at all temperatures. Translating it into a bound on Lyapunov exponent of OTOC, we find a new bound on chaos
\bea
\lambda_{\rm OTOC}\leq {2\pi T\over 1+2 T \bar{\beta}(T)},
\eea
where $\bar{\beta}$ is such that  $C_T(t)$ is  analytic inside the strip $|\Im(t)|\leq \bar{\beta}(T)$. For local systems we find $\bar{
\beta}(T)\geq 2\beta^*$ with $\beta^*$ given by \eqref{betastar} for all $T$. We illustrate this bound for SYK model in section \ref{sec:L}, see Fig.~\ref{Fig:SYK}.

We conclude with a  discussion in section \ref{sec:summary}.

\section{Bound on operator norm growth in Euclidean time}
\label{sec:normbound}
Our goal  in this section is to bound the infinity norm of a local operator evolved in Euclidean time 
\bea
A(-i\beta)=e^{\beta H}A\, e^{-\beta H}, \quad |A(-i\beta)|\leq |A| f(\beta). 
\eea
Here $f(\beta)$ is a bound which depends on the inverse temperature $\beta$, the strength of local coupling $J$ and geometrical properties of the underline lattice model. We argue that our bound \eqref{answer1} (for 1D systems) and \eqref{answer} (for higher dimensions) is optimal for the class of models characterized by the same strength of  the local coupling constant $J$ and lattice geometry encoded in the 
Klarner's constant $\lambda$ and animal histories constant $\varepsilon$ which we introduce later in this work. 

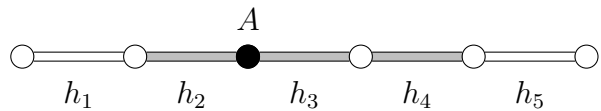
\begin{figure}
\begin{tikzpicture}

\foreach \s in {1,5}
{
\draw[double,double distance = 3pt] ({1.5*(\s - 1)+0.15},0) -- ({-0.15+1.5*(\s)},0);
}

\foreach \s in {2,3,4}
{
\draw[double distance = 3pt,double=lightgray] ({1.5*(\s - 1)+0.15},0) -- ({-0.15+1.5*(\s)},0);
}

\foreach \s in {1,...,6}
{
  \node[draw, circle,radius=5pt] at ({1.5*(\s - 1)},0) {};
}

 \node[draw,fill, circle,radius=5pt] at ({1.5*(3 - 1)},0) {};
\node[] at ({1.5*(3 - 1)},0.5) {\large $A$};

\foreach \s in {1,...,5}
{
  \node[] at ({1.5*(\s - 1)+0.75},-0.5) {\large $h_{\s}$};
}

\end{tikzpicture}
\caption{One-dimensional lattice with short-range interactions Hamiltonian $H=\sum_{I=1}^5 h_I$.
Local operator $A$ sits at a third site counting from the left, between second and third bonds. Bonds highlited in gray form a lattice animal $I=2,3,4$. 
}
\label{Fig:1}
\end{figure}

For simplicity, we first consider nearest neighbor  interaction Hamiltonian in 1D
\bea
\label{sumh}
H=\sum_{I=1}^L h_I,
\eea
where each $h_I$ acts on sites $I$ and $I+1$ and for all $|h_I|\leq J$ for some $J$  \footnote{Time-evolved $A(t)$ will not change if any of the local Hamiltonians $h_I$ is shifted by a constant. Therefore we define $h_I$ such that the absolute value of its largest and smallest eigenvalues are the same}. Any nearest neighbor interaction spin chain would be an example. The operator $A$ will be an one-site operator. An example with $L+1=6$ sites is shown in Fig.~\ref{Fig:1}.

Euclidean time-evolved $A(-i\beta)$ can be expanded in Taylor series 
\bea
\label{Taylor}
A(-i\beta)=A+\beta[H,A]+{\beta^2\over 2}[H,[H,\beta]]+\dots
\eea
Using decomposition \eqref{sumh} operator  $A(-i\beta)$ can be represented as a sum of nested commutators of the form 
\bea
A(-i\beta)=A+\sum_{k=1}^\infty \sum_{\{I_1,\dots,I_k\}} [h_{I_k},[\dots,[h_{I_1},A]]] {\beta^k\over k!}.
\label{nestedc}
\eea
Here the sum is over all sets of indexes $\{I_1,\dots,I_k\}$ which satisfy the following ``adjacency'' condition: first index $I_1$ must be adjacent to the site of $A$, $I_2$ must be adjacent to the endpoints of $I_1$ (which include the site of $A$), $I_3$ is adjacent to the endpoints of the union of $I_1,I_2$, etc. In other words, any subset of bonds $I_1,I_2,\dots,I_\ell$ for $\ell\leq k$ defines a connected cluster. Otherwise, the commutator in \eqref{nestedc} vanishes.

A connected cluster of bonds of any particular shape is called a bond lattice animal. In 1D, all lattice animals consisting of $j$ bonds are easy to classify: they are strings of consecutive bonds from some $I$ to $I+j-1$. In higher dimensions, the number of different bond lattice animals consisting of $j$ bonds  grows  quickly with $j$. 

Each set $\{I_1,\dots,I_k\}$ in \eqref{nestedc} defines a lattice animal, but the same animal may correspond to different sets. This is because indexes can repeat and appear in different orders, subject to the constraints outlined above. If we think of the set $\{I_1,\dots,I_k\}$ as a ``word'' written in terms of ``letters'' $I_\ell$, then corresponding lattice animal defines the alphabet. 

There is a more nuanced characteristics of index sets from \eqref{nestedc}, the order in which new  indexes appear. Namely, we take a set $\{I_1,\dots,I_k\}$ and while going from left to write remove indexes which have already appeared. In this way we obtain a new (shorter) set which also satisfies the  adjacency condition. A particular order is called ``history.'' For example, two sets $\{2,3,2,4,3\}$ and $\{3,3,4,2,4\}$ define the same lattice animal consisting of bonds $I=2,3,4$ but different histories, $\{2,3,4\}$ and $\{3,4,2\}$ correspondingly, see Fig.~\ref{Fig:1}. 

Going back to the sum \eqref{nestedc}, to bound the infinity norm of $A(-i\beta)$ we can bound each nested commutator by $(2J)^k|A|$. Then 
\bea
|A(-i\beta)|&\leq& |A|f(\beta),\\
f(\beta)&=&\left(1+\sum_{k=1}^\infty \sum_{\{I_1,\dots,I_k\}} {(2J|\beta|)^k\over k!}\right),
\label{boundsum}
\eea
and the non-trivial task is to calculate the number of sets $\{I_1,\dots,I_k\}$ for any given $k$, which satisfy the adjacency condition. Evaluating sum \eqref{boundsum}  can be split into two major steps. First step is to calculate the total number $\phi(j)$ of animal histories associated with all possible lattice animals consisting of $j$ bonds.
Second step is to calculate the sum over sets $\{I_1,\dots,I_k\}$ associated with any given history $\{J_1,\dots,J_j\}$. 

This last problem can be solved exactly in full generality. Let's assume we are given a history - a set $\{J\}=\{J_1,\dots,J_j\}$ which satisfies the adjacency condition. We want to know the number of different sets $\{I\}=\{I_1,\dots,I_k\}$ for $k\geq j$ satisfying the adjacency condition such that $\{J\}$ is the history of $\{I\}$. We denote this number by $S(k,j)$. An important observation here is that any given set $\{I\}$ defines a partition of $\{1,2,\dots, k\}$ into $j$ groups  labeled by elements from $\{J\}$ by assigning each number $1\leq i \leq k$ to a group  specified by $I_i$. And vice verse, each partition of $\{1,2,\dots, k\}$ into $j$ groups defines a proper set $\{I\}$ satisfying the adjacency condition. To see that we need to assign each group a unique label from $\{J\}$. We do it iteratively. The element $1$ belongs to a group, which will be assigned the label $J_1$.  Then we consider element $2$. If it belongs to the same group labeled by $J_1$ we move on to element $3$, otherwise we assign the group it belongs label $J_2$. Then we consider elements $3$, $4$ and so on.  In this way all $j$ groups will by labeled by the unique  elements from $\{J\}$ such that  the adjacency condition is satisfied.


In other words, we have established a one-to-one correspondence between the space of proper sets $\{I\}$ for the given history $\{J\}$ with the space of partitions of $k$ elements into $j$ groups. The number $S(k,j)$ of such partitions is the Stirling numbers of the second kind which admits the following representation \cite{abramovich1964handbook}
\bea
\label{S2}
S(k,j)=\sum_{s=1}^j {(-1)^{j-s} s^{k-1}\over (j-s)!(s-1)!}. \label{n}
\eea

If we introduce the number of proper sets $\{I_1,\dots,I_k\}$ in \eqref{boundsum} consisting of $k$ bonds by ${\mathcal N}(k)$, such that 
\bea
\label{k-expansion}
f(\beta)=1+\sum_{k=1}^\infty {\mathcal N}(k){(2J|\beta|)^k \over k!},
\eea
then ${\mathcal N}(k)$ and $\phi(j)$ are related by the Stirling transform,
\bea
\label{Nphi}
{\mathcal N}(k)=\sum_{j=1}^k  S(k,j) \phi(j).
\eea
The inverse relation is $\phi(j)=\sum_{k=1}^j s(j,k) {\mathcal N}(k)$, where $s(j,k)$ are the Stirling numbers of the first kind.  From here in full generality follows \cite{bernstein1995some}
\bea\label{ourresult}
f(\beta)=1+
\sum_{j=1}^\infty \phi(j) {q^j\over j!},
\eea
where
\bea
\label{qdef}
q:=\left(e^{2|\beta| J}-1\right).
\eea
We will derive this identity below 

The expansion in $q$ \eqref{ourresult} has an obvious advantage over \eqref{k-expansion}. Locality is implicit in \eqref{k-expansion}, where the terms at the  order $\beta^k$ come from the lattice animals of all sizes. At the same time \eqref{ourresult} makes locality manifest, terms at the order $q^j$ come only from the lattice animals which have at least $j$ bonds. This representation therefore can be used to establish Euclidean version of Lieb-Robinson bound, see section \ref{sec:LB}.

To evaluate  \eqref{ourresult} we still need to know the number of lattice 
animal histories $\phi(j)$ for a given $j$. In case of 1D systems, those can be calculated exactly, while in higher dimensions we propose an asymptotically tight bound. Hence, we consider these cases separately. 

\subsection{1D systems}
\label{1d}
In one dimension, all lattice animals consisting of $j$ bonds are simply the strings of $j$ consecutive bonds. There are $N(j)=j+1$ such 
animals which include the site of the operator $A$. A convenient way to enumerate them is to count the number of bonds $j_1$ and $j_2$, $j_1+j_2=j$, to the left and to the right of $A$, respectively. 
For the given $j_1,j_2$ there is, obviously, only one animal, $N(j_1,j_2)=1$.

For any given $j_1,j_2$ we denote by $h(j_1,j_2)$ the number of histories associated with this animal, i.e.~the number of different sets $\{J\}=\{J_1,\dots,J_j\}$ such that each $J_i$ belongs to the animal, all $J_i$ in the set are unique and $\{J\}$ satisfies the adjacency condition. Each history $\{J\}$ can be completely parametrized by the order in which the cluster ``grew'' in left and right directions, for example histories $\{2,3,4\}$ and $\{3,4,2\}$ from Fig.~\ref{Fig:1} can be parametrized as ``left,right,right'' and ``right,right,left'' correspondingly. In other words histories with given $j_1,j_2$ are in one to one correspondence with strings of $j$ elements, each element being either ``left'' or ``right,'' and there are in total $j_1$ and $j_2$ elements of each kind. Obviously, the total number of such strings is 
\bea
h(j_1,j_2)={(j_1+j_2)!\over j_1!\, j_2!}.
\eea

Combining all ingredients together, we find the number of lattice histories for all lattice animals of size $j$
\bea
\label{phij1j2}
\phi(j_1,j_2)&=&N(j_1,j_2)h(j_1,j_2)={(j_1+j_2)!\over j_1!\, j_2!},\\
\phi(j)&=&\sum_{j_1+j_2=j} \phi(j_1,j_2)={2^j\over j!}.  \label{phi1D}
\eea
 from \eqref{k-expansion} and \eqref{Nphi} we find in full generality
\bea
\label{f-from-phi}
f(\beta)=1+\sum_{j \geq 1}^\infty\, \sum_{k=j}^\infty \phi(j) S(k,j) {(2|\beta| J)^k\over k!}.
\eea
By definition $k\geq j$. Crucially, expression \eqref{n} vanishes for $1\leq k<j$. Therefore the sum over $k$ can be extended to go from $k=1$ and can be easily evaluated,
\bea
\nonumber
f(\beta)=1+\sum_{j\geq 1}^\infty \sum_{s=1}^{j}  {(-1)^{j-s} \phi(j) \over (j-s)!(s-1)!}{e^{2 \beta J s}-1\over s}.
\eea 
The sum over $s$ can be evaluated explicitly, yielding \eqref{ourresult} \footnote{As a side note, that evaluation of \eqref{f-from-phi} in section \ref{1d} imply Lemma 5 of \cite{Kliesch_2014}. 
Let us consider a fixed lattice animal consisting of $j$ bonds, listed in some arbitrary order $\{J_1,\dots,J_j\}$. One may want to calculate  $G=\sum_{k\geq j}\sum_{\{I_1,\dots, I_k\}}(2J|\beta|)^k/k!$, where the sum is over all sets $\{I_1,\dots, I_k\}$, where each $I_i$ belongs to the set  $\{J_1,\dots,J_j\}$, and each $J_i$ appears in the set $\{I_1,\dots, I_k\}$ at least once.  This is a simplified version of our main calculation, with the adjacency condition being ignored. It is the sum evaluated in Lemma 5 of \cite{Kliesch_2014}.  By
taking a set  $\{I_1,\dots, I_k\}$ from the sum we can associate to it a set $\{I_1,I_{i_2},\dots I_{i_j}\}$ by going from the left to the right and removing repeating labels. As a set  (i.e.~ignoring the order) $\{I_1,I_{i_2},\dots I_{i_j}\}$ coincides with $\{J_1,\dots,J_j\}$. The key point here is the same, the number of sets $\{I_1,\dots, I_k\}$
 associated with the same set $\{I_1,I_{i_2},\dots I_{i_j}\}$ is equal to $S(k,j)$.
If we now sum over all sets $\{I_1,\dots, I_k\}$ associated with a particular $\{I_1,I_{i_2},\dots I_{i_j}\}$, this is exactly the sum evaluated in  \eqref{f-from-phi} with $\phi(j)=1$. Since there are $j!$ different permutations of labels in $\{J_1,\dots,J_j\}$, and thus $j!$ sets  $\{I_1,I_{i_2},\dots I_{i_j}\}$ we therefore obtain $G=q^j$}. Using the explicit value of $\phi(j)$ \eqref{phi1D} we find
\bea
\label{answer1}
f(\beta)=\sum_{j=0}^\infty f(j,\beta)=e^{2q}, \quad f(j,\beta)={(2q)^{j}\over j!}.
\eea
Here $f(j,\beta)$ is a contribution to the bound coming from the clusters which include at least $j$ bonds. 

This result can be further refined. In \eqref{phij1j2} we introduced the number of lattice histories associated with the lattice animal which consists of $j_1$ bonds to the left of $A$, and $j_2$ bonds to the right. Repeating the summation  in \eqref{f-from-phi} we readily find 
\bea
f(\beta)=\sum_{j_1, j_2\geq 0}^\infty f(j_1,j_2,\beta),\quad  f(j_1,j_2,\beta)={ q^{j_1+j_2}\over j_1!\, j_2!}.\ \ \ 
\eea
Here $f(j_1,j_2,\beta)$ is the bound on the norm of the part of $A(-i\beta)$ supported on the cluster of  size $j_1+j_2$. It therefore can be used to obtain the bound in the case of finite 1D lattice, or an infinite 1D lattice with a boundary.

By re-expanding \eqref{answer1} intro Taylor series in $\beta$,
\bea
f(\beta)=\sum_{k=0}^\infty {B_k(2)\over k!} (2J|\beta|)^k,
\eea 
where $B_k$ are Bell polynomials, we find a bound on the norm of individual nested commutators,
\bea
\label{Bell}
| \underbrace{[H,[\dots,[H,A]]]}_{k\ \rm commutators} |\leq |A|\, B_k(2) (2J)^k.
\eea

\subsection{Bethe lattices}
The behavior of $f(\beta)$ differs drastically in one and higher dimensions. To better understand this difference we consider an ``intermediate'' scenario of a short range Hamiltonian define on a Bethe lattice
of  coordination number $z$ \cite{PhysRevB.30.391}. Namely, we assume that each $h_I$ from \eqref{sumh} ``lives'' on a bond and acts on the Hilbert spaces associated with two vertexes adjacent to that bond. 
For any finite $k$ in the Taylor series expansion \eqref{nestedc}  only finite number of bounds are involved  and the corresponding lattice animals (clusters) live on the Cayley tree. Thus, similarly to 1D, there are no loops, but the total number of lattice animals consisting of $j$ bonds grows exponentially, $N(j)\sim \lambda(z)^j$,
\bea
\ln\lambda(z)=(z-1)\log(z-1)-(z-2)\log(z-2).
\eea  
This exponential growth is typical for lattices in higher dimensions $D>1$.

The total number of lattice animal histories $\phi(j)$ can be calculated exactly in this case (see appendix \ref{appx:B}),
\bea
\label{phi}
\phi(j)=(z-2)^j {\Gamma(j+z/(z-2))\over \Gamma(z/(z-2))},
\eea 
leading to the bound 
\bea
\label{Betheanswer}
f(\beta)= (1-(z-2)q)^{-{z/(z-2)}}.
\eea
In other words, the total number of histories $\phi(j)$ grows as a factorial. The same qualitative behavior applies for all higher dimensional lattices. 

As a final remark, we notice that taking $z\rightarrow 2$ in \eqref{Betheanswer} yields $f(\beta)=e^{2q}$, in full agreement with \eqref{answer1}.


\subsection{Higher dimensional systems}
\label{sec:hd}
%
%

The calculations of previous sections can in principle be extended to an arbitrary lattice system, but the number of lattice animal histories is difficult to evaluate exactly. Nevertheless it is known  that the number of different lattice animals $N(j)$ consisting of $j$  bonds (which include a particular site) grows rapidly in higher dimensions. While the exact formula is not known, the asymptotic growth is known to be exponential, and is controlled by the so-called  Klarner's constant $\lambda$,
\bea
N(j) \sim \lambda^j. \label{Klarner}
\eea
By introducing a sufficiently large but $j$-independent constant $C$ we can uniformly bound the number of lattice animals consisting of $j$ bonds by \footnote{To account for a polynomial pre-exponential factor, coefficient $\lambda$ in \eqref{animalgrowth} may need to be taken strictly larger than the Klarner's constant $\lambda$ in \eqref{Klarner}} 
\bea
\label{animalgrowth}
N(j)\leq C\, \lambda^j.
\eea

The number of histories  for any given animal is the number of different sets $\{J_1,\dots,J_j\}$ where all indexes are distinct, subject to the adjacency condition. Let us denote by $h(j)$ the average number of histories for all animals consisting of $j$ bonds. Then, it is trivially bounded by $h(j)\leq j!$. It can be shown that for sufficiently large $j$ \cite{Bouch}
\bea
h(j)\geq {j!\over a^j},
\eea
for some $a>1$. We, therefore, conjecture that for higher dimensional lattices  $h(j)$ is uniformly bounded by 
\bea
h(j) \leq C'\, {j!\over \varepsilon^j},
\label{Evdokiya}
\eea
for some $\varepsilon>1$ and a $j$-independent constant $C'\geq 1$. This bound is trivially satisfied for $\varepsilon=1$. The non-trivial part here is the expectation that \eqref{Evdokiya} correctly captures the leading (exponential) asymptotic behavior of $h(j)$ with some  $\varepsilon>1$, i.e.~\eqref{Evdokiya} is the optimal bound which can not be further improved (excluding polynomial pre-factors). We therefore  introduce here the constant $\varepsilon$ which we call  {\it animal histories} constant and conjecture that it is strictly larger than $1$. In the end of this section we also derive a lower bound on $\varepsilon/\lambda$. By combining \eqref{animalgrowth} together with \eqref{Evdokiya}  
\bea
\label{ee}
\phi(j) = N(j) h(j)\leq C'(\lambda/\varepsilon)^jj!,
\eea
we find the bound 
\bea\nonumber
f(\beta)=\sum_{j=0}^\infty f(j,\beta),\quad f(j,\beta)=C'{\left(q/q_0\right)^{j}},
\eea
Here $f(\beta)$ is defined to be larger than the sum in \eqref{boundsum}.
The coefficient 
\bea
\label{q0}
q_0= {\varepsilon\over \lambda}. 
\eea
characterizes  lattice geometry. Unlike in 1D, where \eqref{answer1} has an additional factorial suppression factor, $f(j,\beta)$ in higher dimensions grows exponentially for sufficiently large $\beta$. Summing over $j$ yields 
\bea
\label{answer}
f(\beta)={C' \over 1-q/q_0}.
\eea 
In contrast to 1D, while  \eqref{answer1} is finite for all $\beta$, \eqref{answer} is finite only for 
\bea
\label{betastar}
|\beta|<\beta^*\equiv \ln(1+q_0)/(2J).
\eea
While \eqref{answer} is only a bound on $f(\beta)$ defined in \eqref{boundsum}, location of the singularity in both cases is the same because it is only sensitive to the asymptotic behavior of $N(j)$ and $h(j)$.

Expanding \eqref{answer} in Taylor series
\bea
f(\beta)=C'\sum_{k=0}^\infty {P_k(q_0^{-1})\over k!}(2J|\beta|)^k,
\eea 
where $P_k$ are the polynomials defined via
\bea
\label{Pk}
P_k(x)={1\over 1+x} \left(x(1+x){\partial \over \partial x}\right)^k(1+x),
\eea
yields a bound on individual nested commutators 
\bea
\label{nested}
| \underbrace{[H,[\dots,[H,A]]]}_{k\ \rm commutators} |\leq |A|C'\, {P_k(q_0^{-1})} (2J)^k.
\eea
The divergence of bound \eqref{answer} at $|\beta|=\beta^*$ is not  an artifact of an overly conservative  counting, as confirmed by a 2D model introduced in \cite{Bouch}, for which $|A(-i\beta)|$ is known to diverge.
We will argue in section \ref{sec:S} that the growth outlined by the bounds (\ref{answer1},\ref{answer}) reflects actual growth of $|A(-i\beta)|$ in non-integrable systems and singularity of \eqref{answer} at finite $\beta$ is a sign of chaos. 
We also note that in case of 1D systems the bound \eqref{answer1} ensures that the operator norm of $A(t)$ remains bounded for any complex $t$.  This is consistent with analyticity of correlation functions in 1D \cite{araki1969gibbs}. On the contrary, in higher dimensions, physical observables may not be analytic.  
We discuss   the relation between  the singularity of $|A(-i\beta)|$ and non-analyticity of physical observables due to a phase transition in section \ref{sec:S} and show that they have different origin.

It is interesting to compare our result for a general lattice in $D>2$ with the exact result for Bethe lattices obtained in the previous section. From  \eqref{phi} and \eqref{ee} we obtain  lattices animal histories constant $\varepsilon$ for Bethe lattices,
\bea
\varepsilon=\left({z-1\over z-2}\right)^{z-1}, \quad  q_0={\varepsilon\over \lambda}={1\over z-2}.
\eea
For any $z\geq 2$,  $\varepsilon >1$  supporting our conjecture that $\varepsilon$ is always strictly larger than $1$. 
Our universal expression \eqref{answer} bounds the exact result \eqref{Betheanswer} from above with any $q_0<1/(z-2)$ and sufficiently large $C'$. 

Bethe lattices provide a lower bound on the combination $q_0={\varepsilon\over \lambda}$ and hence on the critical value $\beta^*$. 
We show in the Appendix \ref{loop} that for any lattice of coordination number $z$, such that each vertex is attached to at most $z$ bonds the number of lattice animal histories is bounded by  $\phi(j)\leq (z-2)^j {\Gamma(j+z/(z-2))\over \Gamma(z/(z-2))}$. We therefore find in full generality
\bea
\label{Bethebound}
  q_0={\varepsilon\over \lambda} \geq {1\over z-2}.
\eea
This bound is stronger than any previously known, as we explain below. 

To conclude this section, we demonstrate the advantage of counting lattice animal histories as is done in \eqref{ourresult} over previously explored approaches.  There is a straightforward way to estimate the number of sets ${I_1,\dots, I_k}$ in \eqref{boundsum} from above by counting the number of ways  a new bond can be added to the set at each step. Provided the lattice has coordination number $z$, starting from the site of $A$, there are $z$ ways to choose $I_1$, at most $z(2z)$ ways to choose $I_2$, $z(2z)(3z)$ ways to choose $I_3$ and so on.  As a result we would get an estimate for $f(\beta)$,
\bea
f(\beta)\leq f_{\rm approx}=\sum_{k=0}^\infty (2J|\beta|)^k z^k = {1\over 1-2J |\beta|z}.
\label{naive}
\eea
This result  was previously obtained in \cite{Abanin_2015,arad2016connecting}.
This gives the following estimate for the location of the pole 
\bea
\label{naivebetastar}
|\beta|={z^{-1}\over 2J}. \label{naivesin}
\eea
The approximation \eqref{naivebetastar} is naive as it overcounts the number of sets $\{I_1,\dots,I_k\}$ assuming the underlying cluster is always of size $k$. We therefore expect  \eqref{naive} to be weaker than our \eqref{answer}, $f(\beta)\leq f_{\rm approx}(\beta)$,  and in particular the location of the singularity \eqref{naivesin} to be smaller than  $\beta^*$ defined in \eqref{betastar}. This can be written  as an inequality
\bea
\varepsilon/\lambda\geq e^{1/z}-1, \label{ah}
\eea
which is  indeed satisfied due to \eqref{Bethebound}. The advantage of \eqref{Bethebound}  becomes apparent in the limit $z\rightarrow 2$ when $\beta^*$ becomes infinite while  \eqref{naivebetastar} remains finite. 


A result  analogous to  \eqref{answer} has been previously established in  \cite{de_Oliveira_2018}, but  importantly there $q_0$ was just inverse of the lattice animal constant, i.e.~Klarner's  constant introduced in previous section,  $q_0=\lambda^{-1}$. Crucially, we improve this result  to account for proper lattice animal histories by introducing $\varepsilon>1$ in \eqref{q0}. Without $\varepsilon$ critical value of $\beta$ where $f(\beta)$ diverges is given by 
$q_0=e^{2J\beta}-1=\lambda^{-1} $ and e.g.~for a cubic lattice in $D$ dimensions $\lambda$ asymptotes to $2De$ when $D\rightarrow \infty$ \cite{de_Oliveira_2018,miranda2011growth}. This value is smaller than  \eqref{naivebetastar} with $z=2D$, meaning  the inequality  \eqref{ah} is not satisfied. To conclude, without taking  lattice animal histories into account, even exact value of $\lambda$ results in a less stringent  bound than \eqref{naivebetastar}, while our bound is always stronger than that due to \eqref{Bethebound}.

\section{Spatial growth in Euclidean time}
\label{sec:LB}
While deriving the bound on the norm of local operators evolved in Euclidean time, \eqref{answer1} and \eqref{answer}, we obtained a stronger result -- a bound $f(j,\beta)$ (or $f(j_1,j_2,\beta)$ in 1D) on spatial growth of $A(-i\beta)$. It can be immediately translated into the Euclidean analog of the Lieb-Robinson bound \cite{LR} on the norm of the commutator of two spatially separated local operators. If $B$ is an operator with finite support located   distance $\ell$  away from $A$ (measured in the Manhattan norm in case of a cubic lattice), then in $D\geq 2$
\bea
\left|[A(i\beta),B]\right|\leq 2|A||B| \sum_{j=\ell}^\infty f(j,\beta)=2|A||B| {C'\, (q/q_0)^{\ell}\over 1-(q/q_0)}, \nonumber
\eea
where we  assumed that $|\beta|<\beta^*$. For  larger $|\beta|$  there is no bound as the sum does not converge. This result means that the local operator can spread to the whole system, no matter how large or even infinite that is, in finite Euclidean time $\beta=\beta^*$. We will argue in section \ref{sec:S} that this is the true physical behavior in the chaotic case and therefore the bound can not be improved to get rid of the divergence at $|\beta|=\beta^*$ in full generality.

In 1D the situation is very different. Assuming local operator $B$ is located $\ell$ bonds away from $A$ we find 
\bea
\nonumber
\left|[A(i\beta),B]\right|&\leq& 2|A||B|  \sum_{j_1=0,j_2=\ell}^\infty f(j_1,j_2,\beta)=\\ &&2|A||B| {e^{2q}\over (\ell-1)!}\int_0^q e^{-t} t^{\ell-1}dt.  \label{LR}
\eea
(If the system is infinite only in one direction  and $A$ is sitting at the boundary, one factor  of $e^{q}$ should be removed.)
Qualitatively the RHS of \eqref{LR} behaves as 
\bea
\label{LRsimplify}
\left|[A(-i\beta),B]\right|\leq 2|A||B|\,  {q^\ell \over \ell!}e^{q} ,
\eea
for $\ell \gg q+1$, and asymptotes to $2|A||B| e^{2q}$ for $\ell \ll q+1$. This means a local operator spreads exponentially fast, to distances $\ell \sim e^{2J\beta}$, in Euclidean time $\beta$.

Exponential spreading of operators in 1D seems to be in agreement with the convergence of the Euclidean variational algorithm of \cite{Beach_2019} in logarithmic time. The connection between  Euclidean Lieb-Robinson bound and 
 the convergence time  is intuitive, but difficult to establish rigorously, in particular, because the latter is sensitive to the choice of initial wave-function. For the integrable models, for which the spreading of operators is at most polynomial (see section \ref{sec:S}), convergence time might be even shorter  because of a well-tuned initial wave-function. For the chaotic systems we expect no fine-tuning of the initial state and hence a direct relation between the  convergence time and Euclidean  Lieb-Robinson bound.

Another possibly intriguing connection is with the studies of Thouless times in chaotic Floquet systems without conserved quantities \cite{Chan_2018}. There, it was noticed that in 1D Thouless time is logarithmic in system size (see also  \cite{Gharibyan:2018jrp}), and finite in $D\geq 2$ (see, however, \cite{Bertini:2018wlu}). That is exactly the same behavior as in the case of Euclidean operator spreading. 
One potential interpretation would be that Thouless time can be associated with the slowest Euclidean mode propagating in the system. Under Euclidean time evolution with a time-dependent random Hamiltonian our extension of Lieb-Robinson bound holds. We also surmise  that in this case spatial growth of all quantities, including the slowest, is qualitatively and  outlined by the bound  with some effective $J,q_0$.
When the system in question has a local conserved quantity, the slowest transport mode is diffusive, leading to $L^2$ scaling of Thouless time \cite{friedman2019spectral}. Thus, to compete this picture it would be  necessary to establish that under Euclidean time evolution time necessary for a diffusive mode to travel across the system is the same as in the Minkowski case, i.e~$\beta\sim L^2$, where $L$ is the system size.

Finally, we notice that  the Euclidean analog of the Lieb-Robinson bound in 1D \eqref{LR} looks similar to the conventional Minkowski bound  \cite{chen2019operator}
\bea
\label{LRM}
\left|[A(t),B]\right|\leq 2|A||B|\,  {(2Jt)^\ell \over \ell!},
\eea
with $2Jt$ substituted by $q(\beta)$.

\section{Constraints on matrix elements}
\label{sec:ME}
\subsection{Individual matrix elements}
Constraints on the infinity-norm of $A(i\beta)^\dagger=A(-i\beta)$ provide an upper bound on the magnitude of matrix elements $A_{ij}=\langle E_i|A|E_j\rangle$ in the energy eigenbasis. Starting from
\bea
\label{identity}
A(-i\beta)_{ij}\equiv \langle E_i|A(-i\beta)|E_j\rangle=A_{ij} e^{\beta(E_i-E_j)}
\eea
we find 
\bea
\label{boundSpain}
|A_{ij}|\leq e^{-\beta(E_i-E_j)} |A(-i\beta)|. 
\eea
This inequality holds for any $\beta$ and we therefore can optimize it over $\beta$. Using explicit form of the bound \eqref{answer1} in 1D we find optimal value of $\beta$ to be (without loss of generality we assumed $\omega=E_i-E_j\geq 0$)
\bea
\label{ob}
\beta=\left\{\begin{array}{lr}
\ln\left({\omega\over 4J}\right)/(2J), &  \omega \geq 4J,\\[2pt]
0,  & 4J\geq \omega.
\end{array}
\right.
\eea
This yields 
\bea
\label{meb}
|A_{ij}|\leq |A| \kappa(\omega),\qquad \omega=|E_i-E_i|,
\eea
where 
\bea
\label{bme1}
\kappa(\omega)&\equiv &\left\{\begin{array}{lr}
{\rm exp}\left\{
2\,\tilde{\omega}\left(1-\ln\tilde{\omega}\right)-2
\right\}, &  \tilde{\omega}=\omega/(4J)\geq 1,\\[2pt]
1,  & \tilde{\omega} \leq 1.
\end{array}
\right.\nonumber
\eea
These results shows that in 1D  for large energy difference  $\omega=|E_i-E_i|\gg  J$ off-diagonal matrix elements $A_{ij}$ decay faster than exponential. For $\omega \leq 4J$ the bound trivializes to $|A_{ij}|\leq |A|$.

In higher dimensions the bound on $A_{ij}$ from \eqref{boundSpain} can not be better than exponential. This is because $f(\beta)$ is a monotonically increasing function of $\beta$ which diverges for some $|\beta|=\beta^*$. In particular
\bea
\label{boundonbound}
e^{-\beta \omega}|A(-i\beta)|\geq e^{-\beta^* \omega} |A| 
\eea
for any $\beta$ and $\omega\geq 0$. To find leading exponent we optimize \eqref{boundSpain}
over $\beta$ to find,
\bea
\label{optimalbeta}
\beta={\ln\left({\omega(1+q_0^{-1})\over 2(\omega+J)}\right)\over 2J},
\eea
and $|A_{ij}|\leq |A| \kappa(\omega)$, where $\omega=|E_i-E_i|$,
\bea
\label{bme}
\kappa(\omega)&=& C' q_0^{-1} \tilde{\omega} \left({\tilde{\omega}(1+q_0^{-1})\over 1+\tilde{\omega}}\right)^{-1-\tilde{\omega}},\quad 
\tilde{\omega}=\omega/(2J). \nonumber
\eea
Taking $\omega\rightarrow \infty$ limit, we find that the asymptotic exponential behavior is 
given by  \eqref{boundonbound},
\bea
\kappa(\omega) \lesssim C'' \omega e^{-\beta^*\omega}, \quad \omega\gg J,
\eea
where $C''$ is  some $\omega$-independent constant.

Constraints on individual matrix elements \eqref{bme1} and \eqref{boundonbound}  only depend on energy difference $\omega$. In the case when the system satisfies ETH, off-diagonal matrix elements for $i\neq j$ are known to be exponentially suppressed by the entropy factor, $|A_{ij}|^2\sim  e^{-S}$. Therefore for the chaotic systems the bound will be trivially satisfied unless $\omega$ is extensive.  

The bound analogous to \eqref{bme} has previously appeared  in \cite{de_Oliveira_2018}, with $\beta^*$ given by \eqref{betastar} with $\varepsilon=1$.

\subsection{Constraints on power spectrum}
\label{sec:ps}
Bounds on individual matrix elements found above can be extended to the autocorrelation function of a Hermitian local $A$,
\bea
\label{C}
C(t)\equiv {\rm Tr}(\rho A(t) A), 
\eea
and its power spectrum 
\bea
\nonumber
\Phi(\omega)&=& {1\over 2\pi}\int_{-\infty}^\infty dt\, e^{-i\omega t} C(t)=\\ &&  \sum_{i,j} p_i |A_{ij}|^2\delta (E_i-E_j-\omega). \label{ps}
\eea
Here $\rho$ is an arbitrary density matrix which commutes with the Hamiltonian, $\rho=\sum_i p_i |E_i\rangle \langle E_i|$, ${\rm Tr}\rho=1$. 

Although bounds on moments $M_k$ derived below are universal for all $\rho$, in what follows we will be most interested in the case when  $\rho$ is the Gibbs ensemble $\rho=e^{-H/T}/Z$, in which case autocorrelation function and power spectrum will be denotes by $C_T$ and $\Phi_T$ correspondingly.
As a function of  complex argument $C_T$ satisfies,
\bea
\label{reflsymm}
C_T(t-i/(2T))&=&C_T(-t-i/(2T)),\\ C_T(t^*)&=&(C_T(t))^*.
\eea

First we notice that 
\bea
|C(t)|\leq |A(t/2)|^2\leq |A|^2 f^2(|t|/2),
\eea
for any complex $t$, which guarantees analyticity of $C(t)$ for 1D systems on the entire complex plane. 

Using the bound on individual nested commutators \eqref{Bell} and \eqref{nested} one can bound the growth of Taylor coefficients of $C$,
\bea
\label{momenta}
M_k=\int\limits_{-\infty}^\infty \Phi(\omega)\omega^k d\omega={\rm Tr}(\rho \underbrace{[H,[\dots,[H,A]]]}_{k\ \rm commutators} A).\quad
\eea
To obtain an optimal bound, nested commutators should be split equally between  two $A$'s using cyclicity of trace 
\bea
|M_{2k+i}| &\leq & |A|^2 (2J)^{2k+i} R_k R_{k+i},\quad i=0,1.
\eea
Here $R_k=B_k(2)$ for infinite 1D system, $R_k=B_k$ for semi-infinite 1D system with a boundary, and $R_k=C' P_k(q_0^{-1})$ for $D\geq 2$.
 
Using the asymptotic behavior of Bell polynomials  \cite{khorunzhiy2019asymptotic}
\bea
B_n(x)\sim \left({n\,\big(1+o(1)\big)\over e \log(n/x)}\right)^n,\quad n\gg x,
\eea
and the Stirling approximation formula, the bound on moments for $k\gg 1$ can be rewritten as  (for the infinite 1D system)
\bea
\label{mu1d}
|M_k|\leq |A|^2 (2J)^k  \left({k\over 2\, e \log k}\right)^k \times e^{o(k)}.
\eea
It is easy to see that the Taylor series of $C_T(t)$ converges in the whole complex plane, as was pointed out above.

In higher dimensions, to find asymptotic behavior of $P_k(x)$ for large $k$, we use the following representation
\bea
P_k(x)=\sum_{j=1}^n j!\, S(k,j)\, x^j ={1\over 1+x}\sum_{j=1}^\infty j^k \left({x\over 1+x}\right)^j. \nonumber
\eea
Substituting the sum over $j$ by an integral and taking saddle point approximation gives
\bea
\label{mubound}
|M_k| \leq |A|^2 \left({q_0\over 1+q_0}\right)^2\left({k\over 2 e \beta^*}\right)^k \times e^{o(k)}.
\eea

Focusing on the case when $\rho=e^{-H/T}/Z$, \eqref{mubound} guarantees that Taylor series of $C_T(t)$  converges absolutely  inside the disc $|t|\leq 2\beta^*$. By representing $C_T$ as a sum over individual matrix elements it is easy to see that if the sum for $C_T(-i\beta)$ is absolutely convergent, then it is absolutely convergent for any  $C_T(t)$, $\Im(t)=-i\beta$.
Therefore $C_T(t)$ is analytic inside the strip $2\beta^*>\Im(t)> -2\beta^*$. Because of reflection symmetry \eqref{reflsymm} function $C_T(t)$ must  be analytic inside a wider strip $2\beta^*>\Im(t)> -2\beta^*-1/T$ \footnote{If $1/(2T)\leq 2\beta^*$, a union of an original strip $|\Im(t)|<2\beta^*$ and its reflection around the point $\beta=-1/(2T)$ is a wider strip $2\beta^*>\Im(t)>-2\beta^*-1/T$. Function $C_T(t)$ has to be analytic there. If  $1/(2T)>2\beta^*$ the same union consists of two strips, $2\beta^*>\Im(t)>-2\beta^*$ and $2\beta^*-1/T>\Im(t)>-2\beta^*-1/T$. It is easy to show though that $C_T$ has to be analytic also in between, $-2\beta^*>\Im(t)>2\beta^*-1/T$. From the definition $C_T(t)={\rm Tr}(\rho^a A \rho^b A)$, $a=i t+1/T$, $b=it$, and positivity $\Re(a),\Re(b)>0$ it follows that the sum over Hilbert space converges, $C_T$ is well defined and therefore analytic.}.
Hence symmetrically ordered  autocorrelation function 
\bea
\label{Wight}
C^W_T(t)\equiv {\rm Tr}(\rho^{1/2}A(t)\rho^{1/2}A)=C_T(t-i/(2T)),
\eea
is analytic inside the  strip $2\beta^*+1/(2T)>\Im(t)> -2\beta^*-1/(2T)$, which is wider than the strip of analyticity of $C_T(t)$, and indicates a more rapid exponential decay of the power spectrum $\Phi^W_T$ of \eqref{Wight}
 in comparison with $\Phi_T(\omega)$.

The logic above is general and does not require any specific details of $M_k$.  Using reflection symmetry \eqref{reflsymm} we have shown in full generality that if $C_T(t)$ develops a singularity at $t=\pm i 2\beta^*$, then $C^W_T(t)$ is analytic at least inside the strip $|\Im(t)|\leq 2\beta^* +1/(2T)$. 

There is another integral bound on power spectrum, valid for any density matrix $\rho$ which commutes with  $H$. 
By integrating \eqref{ps}  we find the following inequality
\bea
\nonumber
\int_{\omega}^\infty d\omega' \Phi(\omega')\equiv \sum_{E_i\geq E_j+\omega} p_i\,  |A_{ij}|^2=\qquad \\ \nonumber
\sum_{E_i \geq E_j+\omega}  p_i\,  e^{-2\beta(E_i-E_j)}|A(-i\beta)_{ij}|^2 \leq \\ \nonumber
\sum_{E_i \geq E_j+\omega}  \! \! \! \!\!\! p_i\, e^{-2\beta \omega }|A(-i\beta)_{ij}|^2 \leq 
e^{-2\beta \omega }\sum_{i, j}  p_i\,  |A(-i\beta)_{ij}|^2 =\\ \nonumber e^{-2\beta \omega }\, {\rm Tr}(\rho A(-i\beta)  A(i\beta))\leq e^{-2\beta \omega } |A(-i\beta)|^2.
\eea
Here $\omega$ is non-negative and in the second equality we used \eqref{identity} with an arbitrary positive $\beta$. Now we can use $|A(-i\beta)|\leq |A|f(\beta)$ and optimize over $\beta$, yielding
\bea
\label{abaninbound}
\int_{\omega}^\infty d\omega' \Phi(\omega')\leq |A|^2 \kappa(\omega)^2.
\eea
Function $\kappa$ is given by \eqref{bme1} and \eqref{bme} in $D=1$ and $D\geq 2$ correspondingly.

When $\rho$ is maximally mixed state, i.e.~temperature $T$ is infinite, the bound can be strengthen to 
\bea
\label{ka}
\int\limits_{|\omega'|\geq \omega} d\omega' \Phi(\omega')\leq |A|^2 \kappa(\omega)^2.
\eea

We would like to emphasize that all bounds discussed above, i.e.~bounds on $M_k$ and \eqref{abaninbound}, are integral in form.  We do not know a rigorous way to directly constrain asymptotic behavior of $\Phi(\omega)$. At the same time at physical level of rigor, if we assume that $\Phi(\omega)$ is a smoothly behaving function at large $\omega$, analyticity of $C(t)$ inside the strip $|\Im(t)|<2\beta^*$ immediately implies that power spectrum in $D\geq 2$ is exponentially suppressed by
\bea
|\Phi(\omega)|\lesssim |A|^2 e^{-2\beta^* \omega}, \qquad \omega \rightarrow \infty.
\eea
In 1D we similarly find super-exponential suppression 
\bea
|\Phi(\omega)|\lesssim |A|^2 e^{-\omega(1+\ln(4J/\omega))/J},  \qquad  \omega \rightarrow \infty.
\eea
The bound on moments  for large $k$  (\ref{mu1d},\ref{mubound}) and the integral bound \eqref{abaninbound} for large $\omega$ follow from here via saddle point approximation. 

Superexponential suppression of $\Phi(\omega)$ emphasizes peculiarity of one-dimensional systems. In particular, it implies that high frequency conductivity \cite{Mukerjee_2006} and energy absorption  \cite{Abanin_2015}  for such systems will be superexponentially suppressed. This is a very special behavior, which should be kept in mind in light of  the numerical studies, which  are often limited to one dimensions, and therefore may not capture correct physical behavior.

An exponential bound on the integral of $\Phi(\omega)$ was first established in 
\cite{Abanin_2015}, where the authors also noted superexponential suppression in 1D, albeit without proposing an explicit analytic form.

\section{Finite size scaling and chaos}
\label{sec:S}
The bounds obtained in section \ref{sec:normbound} correctly account for the number of non-trivial nested commutators $[h_{I_k},[\dots,[h_{I_1},A]]]$ but do not take into account peculiarities of individual local Hamiltonians $h_I$. We therefore expect our bound to be strongest possible among the uniform bounds for the entire family of local short-ranged Hamiltonians defined on a particular lattice. We further assumed that each nested commutator  is equal to its maximal possible value $(2J)^k |A|$. This is certainly too conservative, but for the chaotic systems, i.e.~in absence of some additional symmetries, we expect a finite fraction of nested commutators to grow as a power of $k$. We therefore expect that for large $\beta$ our bounds \eqref{answer1} and \eqref{answer} to correctly describe growth of operator norm in local chaotic systems with some effective values of $J$, as it happens in \cite{Bouch}. In particular in one dimensions we expect $|A(-i\beta)|$ to grow double-exponentially, and in higher dimensions we expect $|A(-i\beta)|$ to diverge at some finite $\beta^*$. 

We similarly expect the bound on spatial growth outlined in section \ref{sec:LB} to correctly capture the spread of local operators when the system is chaotic. An indirect evidence to support that comes from the numerical results of \cite{Beach_2019}, i.e.~logarithmic convergence time of a numerical Euclidean time algorithm, in agreement with \eqref{LRsimplify}.

Below we further outline how $|A(-i\beta)|$ reflects chaos of the underlying system   when the system size is finite. 
It follows from \eqref{ourresult} that for large $\beta$ animal histories with the largest number of bonds will dominate, 
\bea
f(\beta)\propto q^j\sim e^{2J j |\beta|}, \label{growth}
\eea
where $j$ is the total number of bonds in the system, i.e. $j$ is proportional to the volume. Let us compare this behavior with the growth of the Frobenius norm,
\bea
C(-i\beta)={{\rm Tr}(A(-i\beta)A)\over {\rm Tr(1)}}=\sum_{ij} e^{\beta(E_i-E_j)}{|A_{ij}|^2\over {\rm Tr(1)}}.
\eea 
At large $\beta$ leading behavior is 
\bea
C(-i\beta) \propto e^{\beta \Delta E} \label{asym},
\eea where $\Delta E$ is the maximal value of $\Delta E=E_i-E_j$ such that corresponding matrix element $A_{ij}$ is not zero. (In other words $\Delta E$ is the support of $\Phi(\omega)$.) For the chaotic systems satisfying Eigenstate Thermalization Hypothesis we expect most matrix elements to be non-zero, even for extensive $\Delta E$, matching extensive behavior  of $2J j$ in \eqref{growth}. 

Assuming qualitative behavior of \eqref{answer} is correct for non-integrable systems, 
going back to thermodynamic limit in $D\geq 2$, we expect a singularity of $|A(-i\beta)|$ and $C(-2i\beta)$ at some finite $\beta$. 
This singularity has a clear interpretation in terms of $A$ spreading in  the operator space. We first interpret $(A|B):={\rm Tr}(A^\dagger B)/{\rm Tr}(1)$ as a scalar product in the space of all operators and denote corresponding Frobenius norm of $A$ by $|A|_F\equiv (A|A)^{1/2}$. Then if $A$ were typical, i.e.~random in the space of all operators, 
\bea
\nonumber
C(-i\beta)={\rm Tr}(A(-i\beta)A)=|A(-i\beta/2)|_F^2 {Z(\beta)Z(-\beta)\over Z(0)^2},\quad \\ Z(\beta)\equiv {\rm Tr}\, e^{-\beta H}.\, \,\, \nonumber
\eea
Euclidean time evolution can be split into two parts, $A(-i(\beta+\beta'))=e^{\beta' H}A(-i\beta)e^{-\beta' H}$ such that 
\bea
\label{toda}
C(-i(\beta+\beta'))=(A(-i\beta/2)|e^{i\,\beta' {\rm adj}_H}|A(-i\beta/2)).
\eea 
At time $\beta=0$ we start with a local operator, which is not typical. In principle $A(-i\beta/2)$ only explores a particular trajectory in the space of all operators, and therefore can not be fully typical at any $\beta$.  Yet, if we assume that by the time $\beta$ the trajectory of $A$ has explored substantial part of operator space such that  $A(-i\beta/2)$ can be considered  typical enough, we obtain 
\bea
C(-i(\beta+\beta'))\approx C(-i\beta) {Z(\beta')Z(-\beta')\over Z(0)^2}. \label{CC}
\eea
Taking into account that free energy  $\ln(Z)$ is extensive, we immediately see that \eqref{CC} diverges for any $\beta'>0$.   Hence, the singularity of $C(-i\beta)$ and thus also of $|A(-i\beta/2)|$ marks the moment when $A(-i\beta/2)$ becomes typical. This picture is further developed in \cite{dymarsky2019toda}, where we show that the singularity of $|A(-i\beta/2)|$ can be associated with delocalization of $A$ in Krylov space. 

It is interesting to note that since $C(t)$ is analytic for local one-dimensional systems, for such systems, even non-integrable,  $A$ never becomes typical and hence these systems can not be regarded as fully chaotic. 

We separately remark that the conventional time evolution $C(t)=(A(t/2)|A(-t/2))$ does not have an interpretation as the Frobenius norm-squared of $A(t/2)$, therefore  \eqref{toda} does not apply and even if $A(t/2)$ becomes sufficiently typical at late $t$, 
the analog of \eqref{CC} may not hold. 

If the system is finite, at large $\beta$ free energy simply becomes $\ln Z(\beta)/Z(0) \sim -\beta E_{\rm m}$, where $E_{\rm m}$ is  extensive  (minimal or maximal ) energy of the system. Hence
\eqref{CC} will be proportional to $e^{\beta' \Delta E}$, where $\Delta E$ is extensive, in full agreement with \eqref{asym}. This gives the following qualitative behavior of $C(-i\beta)$ when the chaotic system is sufficiently large but finite. For small $\beta$, $\ln C(-i\beta)$ will behave as $\propto e^q$ in 1D and $\propto \ln(q_0-q)$ in higher dimensions. This growth will stop at $\beta\sim \log(L)$ in 1D or $\beta\sim \beta^*$ in $D\geq 2$, at which point in both cases $\ln C(-i\beta)$ will be extensive. At later times $\ln C(-i\beta)$ will grow as ${\beta \Delta E}$ with some extensive $\Delta E$. In the non-integrable case the transition between two regimes, ``thermodynamic'' when $C(-i\beta)$ has not yet been affected by the finite system size, and ``asymptotic,'' is very quick, at most double-logarithmic in 1D.

Behavior of chaotic systems described above should be contrasted with integrable models. In this case most matrix elements $A_{ij}$ are zero and for a wide class of systems, including classical spin models and systems with projector Hamiltonians, support of $\Phi(\omega)$ remains bounded in the thermodynamic limit. (In terms of the Lanczos coefficients, introduced in the next section, this is the case of $\lambda=0$.) For such systems the bounds (\ref{answer1}) and (\ref{answer}) will be overly conservative. For sufficiently large systems and  large $\beta$ we expect \eqref{asym}  with  a system size independent $\Delta E$. This asymptotic behavior will emerge in finite system-independent Euclidean time. Infinity norm $|A(-i\beta)|$ will behave similarly. We further can use \eqref{asym} to estimate the Frobenius norm of nested commutators $| \underbrace{[H,[\dots,[H,A]]]}_{k\ \rm commutators} |_F \leq |A|\Delta E^k$. Assuming infinity and Frobenius norms exhibit qualitatively similar behavior we can substantially improve the Euclidean  analog of the Lieb-Robinson bound 
\bea
|[A(-i\beta),B]|\leq 2|A||B| \sum_{k=\ell}^\infty {\beta^k \Delta E^k\over k!} \sim 2|A||B|{\beta^\ell \Delta E^\ell\over \ell !}, \nonumber
\eea
where last step assumes $\beta \Delta E\ll \ell$. This bound has the same structure as the conventional  Lieb-Robinson bound in Minkowski space \eqref{LRM}. Thus, in the  case of non-interacting models or projector Hamiltonians ($\lambda=0$ in the language of next section) we find ballistic spreading of operators for any complex $t$. 

In the case of a general integrable model, the support of $\Phi(\omega)$ is extensive  and the behavior is  more intricate.  In many explicit examples  in the thermodynamic limit $\Phi(\omega)$ decays as a Gaussian, and $C(-i\beta)\propto e^{(J\beta)^2}$ with some appropriate local coupling $J$ \cite{brandt1976exact,perk1977time,liu1990infinite,viswanath2008recursion,calabrese2012quantum}. (This is  the case of $\lambda=1$ in terms of the next section. See appendix \ref{appx:D} where we  derive  the Gaussian behavior starting from $\lambda=1$.) Using the same logic as above this leads to the  Euclidean   Lieb-Robinson bound of the form 
\bea
|[A(-i\beta),B]|\lesssim 2|A||B|{(\beta J)^{2\ell}\over \ell !},
\eea
which indicates a polynomial propagation of the signal $\ell\propto \beta^2$.
For a finite system of linear size $L$ we may expect Gaussian behavior $C(-i\beta)\propto e^{(J\beta)^2}$ up to the times $\beta\propto L^{1/2}$, after which the asymptotic behavior \eqref{asym} should emerge.  Although the model is integrable, $\Delta E$ is extensive, which implies the transition between ``thermodynamic'' and ``asymptotic'' behavior is long and will take up to $\beta \sim L$. This indicates the qualitative difference between integrable and non-integrable (chaotic) models. When the system is finite in both cases the asymptotic behavior is given $C(-i\beta)\propto e^{\beta \Delta E}$ with an extensive $\Delta E$ (except for the $\lambda=0$ case), but asymptotic behavior will emerge quickly, in finite (for $D\geq 2$) or logarithmic (for $D=1$) times in the non-integrable case, while in the integrable case asymptotic behavior will emerge much  slower, after polynomial times in $L$. 

A qualitatively similar picture will also apply if integrability is broken weakly, by a parametrically small coupling. For an operator initially characterized by $\lambda=0$,  the correlation function will first exhibit  \eqref{asym} with some sub-extensive $\Delta E$, which will gradually grow to extensive values. It would be interesting to study this transition in detail, to see if the required times may be parametrically longer than $\beta \sim L$.

We stress that  non-analyticity of $C(t)$ at  imaginary times is due to $A(-i\beta)$ becoming typical and is not related to non-analyticity of free energy  $\ln Z(\beta)$ due to a phase transitions at some  temperature $\beta$. Indeed, $C(t)$ for the SYK model is known to have a pole at imaginary time \cite{MS}, while there is no phase transition and free energy is analytic. On the contrary, for the 3d Ising $\ln Z(\beta)$ is non-analytic due to a phase transition, but $C(t)$ is entire, simply because $A(t)$ explores only a very small part of the corresponding Hilbert space.

In conclusion, we note that the singularity of $|A(-i\beta)|$ and $C(-i\beta)$ at finite $\beta$ in the thermodynamic limit has an IR origin.  A straightforward attempt to extend the analysis of this section to field theoretic systems, which can be obtained from lattice systems via an appropriate limit, fails because both  $|A(-i\beta)|$ and $C(-i\beta)$ are UV-divergent, and this obscures the  IR divergence due to chaos. 
Formulating the criterion of chaos  for QFTs using Euclidean operator growth thus remains an open question. 

\section{Constraints on Lanczos coefficients}
\label{sec:L}
The bound on power spectrum established in section \ref{sec:ps} can be used to constrain  the growth of  Lanczos coefficients. To remind the reader, 
Lanczos coefficients $b_n$ are non-negative real numbers associated with an orthonormal basis in the Krylov space $A_n$ generated by the action of $H$ on a given operator $A_0=A$. Starting from a  scalar product 
\bea
\label{sp}
(A,B)\equiv {\rm Tr}(\rho^{1/2}A^\dagger \rho^{1/2}B),
\eea
and choosing $A$ normalized such that $|A|^2=(A,A)=1$, Lanczos coefficients are fixed iteratively from the condition that operators $A_n$ defined via $A_{n+1}=([H,A_n]-b_n A_{n-1})/b_{n+1}$ are orthonormal, $(A_n,A_m)=\delta_{nm}$.

An autocorrelation function $C^W=(A(t),A)$,  defined via scalar product \eqref{sp},  can be parametrized in a number of ways, via its power spectrum $\Phi^W(\omega)$, Taylor coefficients (moments) $M_k$, or Lanczos coefficients $b_n$. Schematically an asymptotic growth of $b_n$ for large $n\gg 1$ is related to the behavior of $M_k$, $k\gg 1$, high-frequency tail of $\Phi^W(\omega)$, $\omega\rightarrow \infty$, or growth of $C^W(t)$ at the  Euclidean time $t=-i\beta$. But the detailed relation is not always trivial. {\it Assuming}  exponential behavior of power spectrum  at large frequencies
\bea
\label{asps}
\Phi^W(\omega) \sim e^{-(\omega/\omega_0)^{2/ \lambda}},
\eea 
it is trivial to obtain the growth of $M_k$ and $C^W(\beta)$ by calculating corresponding integrals over $\omega$ using saddle point approximation. Although much less trivial, but starting from the power spectrum \eqref{asps}, it is also possible to establish an asymptotic behavior of Lanczos coefficients \cite{lubinsky1993update}
\bea
b_n^2\propto n^\lambda. \label{lambda}
\eea
 The converse relations between asymptotic behavior of $b_n$, $M_k$, $\Phi^W(\omega)$ and $C^W(-i\beta)$ are much more subtle and may not hold. Thus, we show in the appendix \ref{appx:bfromM} that smooth asymptotic behavior of $M_k$ does not imply smooth asymptotic of $b_n$.

It was proposed long ago that $\lambda$ defined in \eqref{lambda} falls into several universality classes, characterizing dynamical systems \cite{liu1990infinite}.  In particular it was observed that $\lambda=0$ for non-interacting and $\lambda=1$ for interacting integrable models. (It should be noted that since $\lambda$ characterizes a particular operator, the same system may exhibit several different values of $\lambda$.)  Recently it was argued in  \cite{Parker_2019} that $\lambda=2$ is a universal behavior in chaotic systems in $D\geq 2$. 

To thoroughly investigate possible implications of this conjecture, it is desirable to derive the constraints on the behavior of  $C^W$, $\Phi^W$, and $M_k$ starting directly from the assumption that $b_n$ is a  smooth function of $n$ for large $n$. 
In full generality Lanczos coefficients  $b_n$ are related to the moments $M_k$ via 
\bea
\label{Dyck}
M_k=\!\!\!\!\sum_{h_1\dots h_{k-1}} b_{(h_0+h_1)/2} b_{(h_1+h_2)/2} \dots b_{(h_{k-1}+h_{k})/2}.\,\,
\eea
Here the sum is over Dyck paths parameterized by the sets satisfying $h_0=h_k=1/2$, and $h_{i+1}=h_i\pm 1$,  $h_i> 0$. {Assuming} \eqref{lambda} out goal is to deduce an  asymptote of $M_{2k}$ using \eqref{Dyck}. We develop the approach of summing over weighted Dyck paths in the appendix \ref{appx:D}. Here we just mention main results. If $b_{n}^2$ is asymptotically a smooth function of $n$, path integral over Dyck paths can be evaluated via saddle point approximation by identifying a trajectory in the space of indexes, which gives the leading contribution. Thus if $b_n$ is smooth, $M_k$ is also smooth.  Furthermore, if $\lambda=2$, $b_{n}^2 \sim \alpha^2 n^2$, and the leading order behavior is
\bea
M_{2k}\approx \left(2\alpha\over \pi\right)^{2k} (2k)!
\label{DyckM}
\eea
Thus,  starting from the asymptotic behavior $b_n^2\propto \alpha^2 n^2$ we  necessarily find that $C^W$ has a singularity at $\beta=\pi/(2\alpha)$, in full agreement with the  conjecture of previous section that singularity in Euclidean time is the characteristic property of chaos.

Provided $C^W(t)$ is analytic inside a strip $\Im(t)\leq \bar{\beta}_W$ for some $\bar{\beta}_W$ would immediately imply a bound
\bea
\label{boundonL}
\alpha\leq {\pi\over 2\bar{\beta}_W}.
\eea
When $\rho\propto e^{-H/T}$, provided autocorrelation function $C_T$   \eqref{C} is analytic inside $|\Im(t)|\leq \bar{\beta}(T)$,  function  $C^W_T$ defined in \eqref{Wight} will b analytic at least inside $|\Im(t)|\leq \bar{\beta}_W=\bar{\beta}(T)+1/(2T)$ (see the discussion in section \ref{sec:ps}) and therefore 
\bea
\label{boundonLT}
\alpha\leq {\pi T \over 1+2 T \bar{\beta}(T)}.
\eea
We have also established  in section \ref{sec:hd} that $\bar{\beta}(T) \geq 2\beta^*$ for all $T$. 

The coefficient $\alpha$ has been recently conjectured to bound 
maximal Lyapunov exponent governing exponential growth of the out of time ordered correlation function (OTOC) \cite{Parker_2019,Murthy:2019fgs}, $\lambda_{\rm OTOC}\leq 2\alpha$. This leads to the improved bound on chaos 
\bea
\lambda_{\rm OTOC}\leq  {2\pi T\over 1+4 T \beta^*},
\label{OTOC}
\eea
which is stronger than the original bound $\lambda_{\rm OTOC}\leq {2\pi T}$ of \cite{MSS}.  In the limit of quantum field theory, $(\beta^*)^{-1}$ will be of the order of UV-cutoff, reducing \eqref{OTOC} to the original bound. Yet the new bound is non-trivial for discrete models exhibiting chaos.

\begin{figure}[t]
\includegraphics[width=0.5\textwidth]{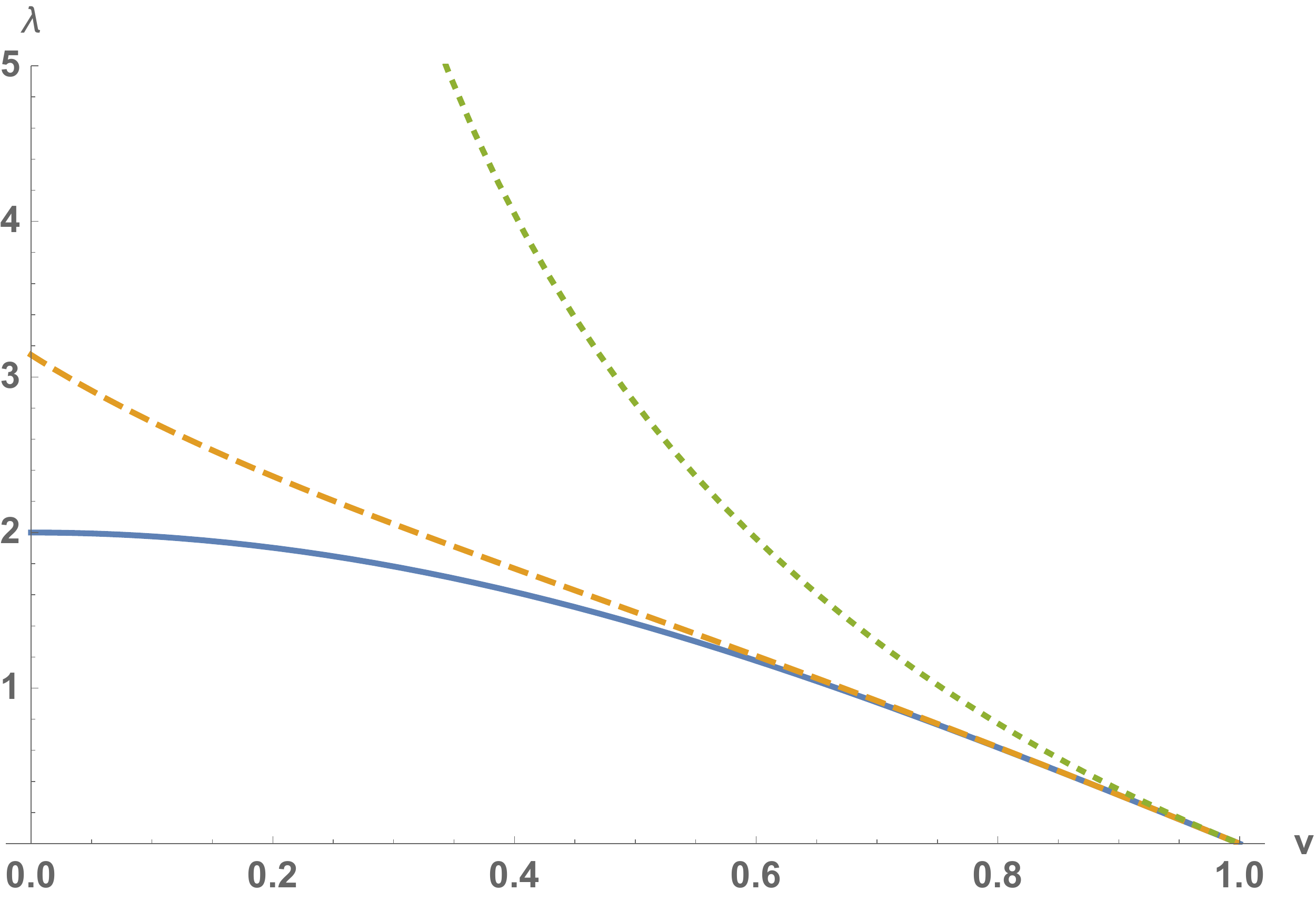}
\caption{Lyapunov exponent $\lambda_{\rm OTOC}$ for the SYK model as a function of parameter $v$, which  is related to temperature, $\pi v T=\cos(\pi v/2)$.
Limit $v\rightarrow 0$ corresponds to high temperatures, $v\rightarrow 1$ to small temperatures.
 Blue line -- exact analytic result \eqref{OTOCexact}, orange  dashed line -- improved bound \eqref{OTOC} with $2\beta^*=1$, 
green dotted  line -- original Maldacena-Shenker-Stanford bound $2\pi T$.}
\label{Fig:SYK}
\end{figure}

To illustrate the improved bound, we plot  \eqref{OTOC} in Fig.~\ref{Fig:SYK}  for the SYK model in the large $q$-limit \footnote{Here $q$ is a parameter of SYK model and should not be mixed with $q(\beta)$ defined in \eqref{qdef}.}
against the exact value of $\lambda_{\rm OTOC}$, evaluated in \cite{MS,Parker_2019}. We take $2\beta^*=1$ to ensure that the autocorrelation function  $C_T$  is analytic inside $\Im(t)<2\beta^*=1$ for all $T$. Temperature $T$ is parametrized via $1\geq v\geq 0$, $\pi v T=\cos(\pi v/2)$ such that the exact Lyapunov exponent is
\bea
\label{OTOCexact}
\lambda_{\rm OTOC}=2\cos(\pi v/2).
\eea

We have emphasized above that for 1D systems with short range interactions $C_T(t)$ has to be analytic in the entire complex plane. This imposes a bound on the growth of Lanczos coefficients. Assuming $b_n$ is a smooth function of $n$  \cite{Parker_2019} proposed that the asymptotic growth in 1D non-integral systems will acquire a logarithmic correction 
\bea
\label{1Dlog}
b_{n+1}\approx \alpha {n\over \log(n/n_0)}.
\eea
Using the integral over weighted Dyck paths in the appendix \ref{appx:D}, we find this to be consistent with the behavior of  $M_k$  outlined in \eqref{mu1d} provided
\bea
\alpha=\pi J/2.
\eea

Sum over Dyck paths in 
the case of $\lambda=1$ associated with integrable systems is discussed  in the appendix \ref{appx:D}. Since for the local models $C^W(t)$ is analytic inside a sufficiently small vicinity of $t=0$, asymptotic behavior with $\lambda>2$ in such systems is excluded.

\section{Conclusions}
\label{sec:summary}

We have derived a number of rigorous bounds on the infinity norm of a local  operator evolved in Euclidean time, and extended them to  autocorrelation function \eqref{Cdef}.   The novel ingredient of our approach is the counting of {\it  lattice animal histories} and formula \eqref{ourresult}, using which we solved exactly combinational problem of counting nested commutators  for Bethe lattices (and establish acorrect asymptotic for lattices in $D\geq 2$). 
Some of  the  bounds  derived in this paper were known before.  We improved numerical coefficients, including  the location of the singularity $\beta^*$ in $D\geq 2$.  Our results are strongest possible among the bounds uniformly valid for all local Hamiltonians characterized by the same $|h_I|\leq J$ defined on a lattice of  a particular geometry. 

We have also established Euclidean version of Lieb-Robinson bound on the spatial operator growth. In 1D operators spread at most exponentially, while in $D\geq 2$ operators can reach spatial infinity in finite Euclidean time. When the system is integrable, in all $D$ operators spread polynomially.  


As a main point of this paper, we advocated that Euclidean operator growth reflects chaos in the underlying quantum system. If the system is chaotic, the norm growth and spatial growth are maximal possible  and the operator norm diverges at some finite Euclidean time. We interpreted this divergence as a consequence of typicality in  Krylov space.

There are several distinct characteristic properties of chaos for many-body quantum systems. One is the Eigenstate Thermalization Hypothesis \cite{srednicki1994chaos}, which is concerned with individual matrix elements. Another popular probe is out of time ordered correlation function, which extends the notion of exponential Lyapunov growth to quantum case.  Its use as a characteristic of many-body quantum chaos was pioneered in  \cite{PhysRevE.89.012923,Elsayed_2015,Tarkhov_2018} and brought  to the spotlight by applications to quantum gravity \cite{MSS}.  Despite recent efforts \cite{Lensky:2018hwa,Foini_2019,Chan:2018fsp,Murthy:2019fgs} there is no clear understanding of how to relate these two characteristics  of chaos to each other. We hope that the Euclidean growth, which on the one hand is related to ETH via the behavior of $C(-i\beta)$ at large $\beta$, see \eqref{asym}, and on the other hand is related to OTOC via the bound \eqref{OTOC}, may provide such a bridge.

\section*{Acknowledgments}

We would like to thank Dima Abanin, Xiangyu Cao, Nick Hunter-Jones, Vadim Oganesyan, and Dan Parker for discussions. We are also grateful to Vladimir Kravtsov,  William Berdanier, and Sarang Gopalakrishnan for raising our interest in Bethe lattices and for discussions. 

AD gratefully acknowledges support from the Simons Center for Geometry and Physics, Stony Brook University at which some of the research for this paper was performed. AD is supported by the National Science Foundation under Grant No. PHY-1720374.

\bibliography{EG}
\newpage

\appendix
\section{Animal histories growth\\ on Bethe lattices}
\label{appx:B}
We consider Bethe lattice of coordination number $z$ and would like to calculate the total number of  lattice animal histories for all possible lattice animals  (clusters) consisting of $j$ bonds. Each lattice animal can be characterized (non-uniquely) by the vector $a_i$ for $i=0,1,\dots,z$ where $a_i$ is the number of vertexes attached to $z-i$ bonds (of that cluster). Sine the considered clusters are connected, either $a_z=1$, in which case $j=0$ and $a_i=0$ for $i<z$, or $a_z=0$. 

Consider any given lattice animal history associated with a lattice animal with a given $a_i$. We can add one additional  bond by attaching it to any vertex, which has less than $z$ bonds already attached to it. If we attach a bond to a vertex with $k$ bonds already attached to it, the new lattice animal (and associated lattice animal history), is charactered by a new set 
\bea
\label{add}
a'_i=a_i+e^{(k)}_i,\quad e^{(k)}_i=\delta_{z-1,i}-\delta_{z-k,i}+\delta_{z-k-1,i}.\quad
\eea
This equation simply reflects the fact that the new cluster has one more vertex with only $1$ bond attached to it, one more vertex with $(k+1)$ bonds attached to it, and 
one less vertex with $k$ attached bonds. 

The total number of bonds $j$ can be expressed through $a_i$ as follows
\bea
\label{ja}
j={\sum_{i=1}^z i a_i-z\over (z-2)}.
\eea
It can be easily checked that adding one bond via \eqref{add} increases $j$ by one, and taking $a_z=1$, $a_i=0$ for $i<z$ gives $j=0$.

We denote the total number of lattice animal histories (associated with all possible animals) consisting of $j$ bonds by $\phi(j)$. (The total number of histories characterized by $a_i$ can be denoted by $\phi(a_i)$. Then $\phi(j)$ is a sum of $\phi(a_i)$ over all possible vectors $a_i$ with non-negative coefficients satisfying \eqref{ja}.)
Given a particular lattice animal history, there are many ways one bond can be added. If we decide to add a bond to vertex which already has $k<z$ bonds attached to it, we will have $a_{z-k}$ vertexes to choose from and $(z-k)$ possibilities for each vertex we chose.  Hence, in total, each lattice animal history parametrized by $a_i$ gives rise to 
\bea
\label{A3}
\sum_{k=0}^{z-1} (z-k) a_{z-k}=j(z-2)+z
\eea
new animal histories consisting of $j+1$ bonds. If we sum over all possible animal histories with $j$ bounds, we  should find the total number of animal histories consisting of $j+1$ bonds, 
\bea
\label{A4}
\phi(j+1)=\phi(j)(j(z-2)+z).
\eea
This immediately yields 
\bea
\label{A5}
\phi(j)=(z-2)^j {\Gamma(j+z/(z-2))\over \Gamma(z/(z-2))},
\eea
where we additionally required $\phi(0)=1$.

While this is not necessary for the bound on operator norm growth, for completeness we derive the number of lattice animals consisting of $j$ bonds, $N(j)$. We first consider all lattice animals which originate at the same vertex and extend into one particular direction (``branch'') on the Bethe lattice. If the number of such animals is $n(j)$, then it must satisfy the recursive relation 
\bea
n(j)=\sum_{j_1+\dots j_{z-1}=j-1} n(j_1)\dots n(j_{z-1}). \label{iter}
\eea
It reflects the fact that we can ``move'' the initial point by one bond inside the branch and decompose $j$ into $j=\sum_{i=1}^z j_i$, $j_z=1$. In \eqref{iter} we also use that $n(1)=1$.  This gives in full generality 
\bea
n(j)={\Gamma((z-1)j+1)\over \Gamma(j+1)\Gamma(2+(z-2)j)}.
\eea
The full number of lattice animals $N(j)$ is related to $n(j)$ via
\bea
N(j)=\sum_{j_1+\dots j_{z-1}+j_z=j} n(j_1)\dots n(j_{z-1})n(j_{z}),
\eea
with the total number being 
\bea
N(j)={z\,\Gamma((z-1)j+z)\over \Gamma(j+1)\Gamma(z+1+(z-2)j)}.
\eea
At large $j$ this number grows as $\lambda^j$ with the Klarner's constant 
\bea
\ln\lambda(z)=(z-1)\log(z-1)-(z-2)\log(z-2).\qquad
\eea 

\section{Animal histories growth on arbitrary lattices in $D\geq 2$} 
\label{loop}
In this section we consider an arbitrary lattice of coordination  number $z$, which means that each vertex is adjacent to at most $z$ bonds. Similarly to previous section, we will characterize a lattice animal (cluster) by a set of numbers $a_i$, $i=0,\dots,z$ where $a_i$ is the number of vertexes attached to $z-i$ bonds of that cluster. From here we can immediately find the total number of bonds,
\bea
j={\sum_{i=0}^z (z-i)a_i\over 2}. \label{jf}
\eea
The main different between general case and the case of Bethe lattices is the possibility of loops.  We define the number of loops $\ell$ of a given lattice animal as the minimal number of bonds which should be removed for the animal to have a tree topology. Then  Euler's characteristic formula  gives $\sum_{i=0}^z a_i -j+\ell=1$.
From here and \eqref{jf} we readily find
\bea
\label{jb}
j={\sum_{i=1}^z i a_i-z(1-\ell)\over (z-2)},
\eea
which is a generalization of \eqref{ja}.

 Let us denote by $n$ total number of ways one can add a bond to a given animal. This number is the total number of bonds adjacent to the animal but not belonging to it. 
 If each vertex had exactly $z$ bonds adjacent to it,  the  sum  $\sum_{k=0}^{z-1}(z-k)a_{z-k}$ counts the number of bonds which can be added to each vertex of the animal. Since some bonds have both ends adjacent to the animal, the sum $\sum_{k=0}^{z-1}(z-k)a_{z-k}$ includes those bonds twice. 
Furthermore, since some vertexes might actually have less than $z$ bonds adjacent to them, $\sum_{k=0}^{z-1}(z-k)a_{z-k}$ provides an upper bound.  
We therefore have an inequality (compare with \eqref{A3})
\bea
\label{B4}
n\leq \sum_{k=0}^{z-1}(z-k)a_{z-k}=j(z-2)+(1-\ell)z.
\eea
Since $\ell\geq 0$, we can conclude that in full generality $n\leq j(z-2)+z$. This expression does not depend on any details of the animal, except its size $j$. We therefore can bound the growth of animal histories for all animals of size $j$, 
\bea
\phi(j+1)\leq \phi(j)(j(z-2)+z),
\eea
from where follows the inequality 
\bea
\phi(j)\leq (z-2)^j {\Gamma(j+z/(z-2))\over \Gamma(z/(z-2))}.
\eea

\section{Integral over weighted \\ Dyck paths} 
\label{appx:D}
In the context of recursion method Lanczos coefficients $b_n$ define tri-diagonal Liouvillian matrix 
\bea
{\mathcal L}=\left(\begin{array}{ccccc}
0 & b_1 & 0 & 0 &\ddots \\
b_1 & 0 & b_2 & 0 & \ddots \\
0 & b_2 & 0 & b3  & \ddots \\
0 & 0 & b_3 & 0 & \ddots  \\
\ddots & \ddots & \ddots & \ddots &\ddots 
\end{array}\right)
\eea
such that correlation function 
\bea
C^W(-i\beta)\equiv (A(-i\beta),A)=\langle 0|e^{{\mathcal L}\beta}|0\rangle, 
\eea
where scalar product of operators is defined in \eqref{sp}, 
and $\langle 0|\dots |0\rangle$ denotes the upper left corner matrix element. By definition, moments $M_k$ are Taylor series coefficients of $C^W$, 
\bea
M_k=\langle 0|{\mathcal L}^k|0\rangle.
\eea
From here and the tri-diagonal form of ${\mathcal L}$ it follows that 
\bea
\label{WDP}
M_{2k}=
\sum_{h_1,\dots ,h_{2k}} \prod_{i=1}^{2k-1} b_{(h_i+h_{i+1})/2},
\eea
while all odd moments vanish (this is also obvious from the symmetry $C^W(t)=C^W(-t)$).
The sum above is over the sets $h_i$ such that $h_1=h_{2k}=1/2$, $h_i>0$, and $h_{i+1}=h_i\pm 1$.

When $k$ is large, sum over Dyck paths becomes a path integral, parametrized by a smooth function $f(t)$, $0\leq t\leq 1$ \cite{Okounkov:2015rir},
\bea
h_i={1\over 2}+2k f(i/(2k)). 
\eea
Function $f(t)$ satisfies
\bea
\nonumber
f(0)=f(1)=0,\quad |f(t_1)-f(t_2)|\leq |t_1-t_2|, \quad f(t)\geq 0. 
\eea
Derivative $f'(t)$ defines an average slope of a ``microscopic'' Dyck path around index $i \approx 2k t$. The path is a sequence of ``up'' and ``down'' jumps with the probabilities $p$ and $1-p$, which vary smoothly, such that $2p(t)-1=f'(t)$. The number of different ``microscopic'' paths ${\mathcal N}[f(t)]$ associated with $f(t)$ is given by the Shannon entropy of $p(t)$, 
\bea
{\mathcal N}[f(t)]\approx e^{S_0}\, \quad S_0=2k \int_0^1 dt\, H(p(t)),\\
 H(p)=-p\log(p)-(1-p)\log(1-p).
\eea
In other words ${\mathcal N}[f(t)]$ is the measure in the path integral over $f(t)$. To verify this result we calculate the total number of Dyck paths, which is known to be given by the Catalan number, 
\bea
{\mathcal C}_k \approx \int {\mathcal D}f(t)\, e^{S_0},
\eea
by evaluating corresponding path integral via saddle point approximation. By interpreting $S_0[f(t)]$ as a classical action, classical EOM is
\bea
{d\over dt} \arctan(f')=0.
\eea
The only solution satisfying boundary conditions is $f(t)=0$, which gives saddle point value 
\bea
{\mathcal C}_k \approx 4^k. 
\eea
This reproduces correct exponential behavior of Catalan numbers, ${\mathcal C}_k\approx 4^k/(k^{3/2}\pi^{1/2})$.

Assuming $b_{n+1}=b(n)$ is a smooth function of index, at least for large $n$, sum over weighted Dyck paths \eqref{WDP} can be represented as an integral 
\bea
\nonumber
M_k&\approx&\int {\mathcal D}f(t)\, e^{S},\\\ S&=&2k \int_0^1 dt \left( H(p(t))+\log b(2k f(t))\right).
\label{actionS}
\eea
In case of asymptotic behavior $b^2(n)=\alpha^2\, n^\lambda$ the EOM is 
\bea
\label{eql}
-\frac{f''(t)}{1-f'(t)^2}=\frac{\lambda }{2 f(t)}.
\eea
For general $\lambda$ this equation can be solved  in terms of an inverse of the Hypergeometric function. We are most interested in two cases, $\lambda=2$ and $\lambda=1$. In the latter case, $b^2(n)=\alpha^2 n$, the saddle point trajectory is $f(t)=t(1-t)$ leading to the asymptotic behavior of moments 
\bea
M_{2k}\approx \left({2k\over e}\right)^k \alpha^{2k}\approx {(2k)!\over k!} \left({\alpha^2 \over 2}\right)^k.
\eea
This gives an exponential growth of $C^W$ at larger $\beta$,
\bea
C^W(-i\beta)\approx e^{(\alpha \beta)^2/2}.
\eea

In the ``chaotic'' case $\lambda=2$ the solution satisfying boundary condition is 
\bea
\label{sinsol}
f(t)={\sin(\pi t)\over \pi},
\eea
and the saddle point value is 
\bea
\label{M2k}
M_{2k}\approx \left({4k \alpha\over e\pi}\right)^{2k}\approx \left({2\alpha\over \pi}\right)^{2k} {(2k)!}
\eea
Provided $C^W$ is analytic inside the strip  $\Im(t)<\bar{\beta}$, the asymptotic growth constant has to be bounded by 
\bea
\label{boundalpha}
\alpha\leq {\pi\over 2\bar{\beta}}.
\eea


Finally we consider the scenario when the growth of Lanczos coefficients acquires logarithmic correction, 
\bea
b(n)= \alpha {n\over \log(n/n_0)}.
\eea
In this case the  action \eqref{actionS} becomes
\bea
\nonumber
S&=&2k\int_0^1 dt \left(H((1+f')/2)+\log(f)-\ln(\ln(2k f/n_0))\right)\\
&+&2k \ln(2k\alpha).
\eea
Taking into account only leading term in $1/\ln(2k/n_0)$ expansion we obtain effective action 
\bea
\label{effaction}
S&=&2k\int_0^1 dt \left(H((1+f')/2)+{\lambda \over 2}\ln(f)\right)+\\
&&2k \ln((2k\alpha)/\ln(2k/n_0)),\quad \lambda=2(1-1/\ln(2k/n_0)).\quad \nonumber
\eea
In other words at leading order the effect of logarithmic correction is in adjusting the scaling parameter $\lambda$.
When $\lambda\approx 2$, the solution of the EOM \eqref{eql} can be found in the power series expansion in $2-\lambda$, with the leading term being simply
\bea
\label{fnew}
f={\sin(\pi t)\over \pi}+O\left({1\over \log(2k/n_0)}\right).
\eea
At leading order the $1/\log(2k)$ correction to $f$ does not affect the on-shell value of \eqref{effaction} evaluated at $\lambda=2$ simply because at leading order \eqref{fnew} is a solution of the EOMs of \eqref{effaction}  with $\lambda=2$. Hence the only correction comes from 
\bea
\delta S=2k\int_0^1 dt\,  \left({\lambda\over 2}-1\right)\log(f),
\eea
where $f$ is given by \eqref{sinsol}. Combining all together we find (compare with \eqref{M2k})
\bea
M_{2k}\approx \left({4k \alpha\over e\pi\ln(2k/n_0)} \right)^{2k} (2\pi)^{2k/\log(2k/n_0)}.
\eea
It is more convenient to work with the logarithm of moments, 
\bea
\nonumber
{\ln M_{2k}\over 2k}= \ln\left({2k \alpha\over e\pi}\right) -\ln \ln(2k/(2\pi n_0)) +o(1/\ln(k)).
\eea 
Comparing this with the asymptotic behavior of moments in 1D \eqref{mu1d}, we identify $\alpha=\pi J/2$, while matching $n_0$ would exceed the available precision of \eqref{mu1d}.

\section{Reconstruction of $b_n$ from $M_k$} 
\label{appx:bfromM}
\begin{figure}[t]
\includegraphics[width=0.5\textwidth]{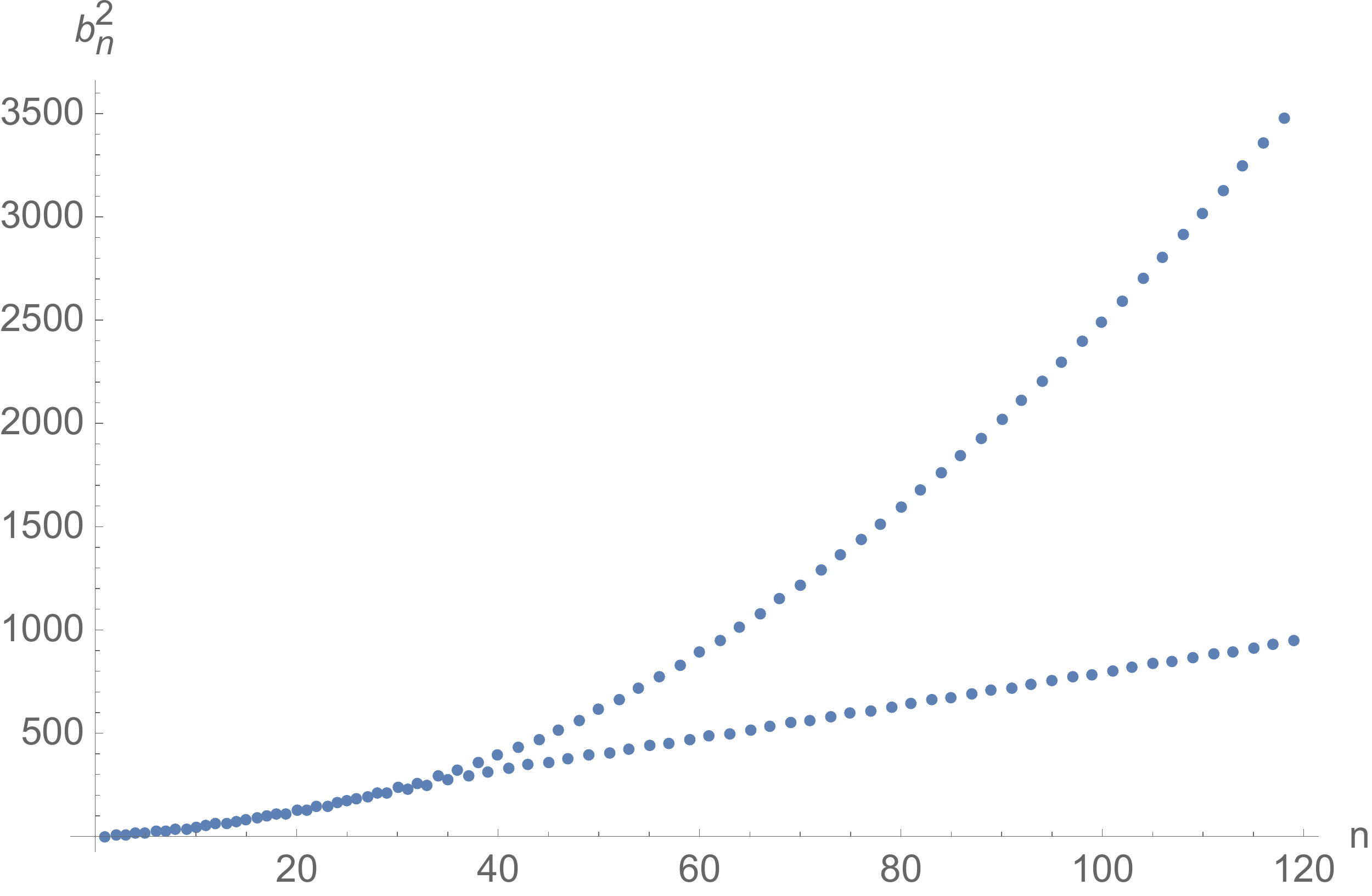}
\caption{Lanczos coefficients $b_n^2$ associated with the moments $M_{2k}=B_{2k}\equiv B_{2k}(1)$. Choosing different $m$ in \eqref{mock} leads to a qualitatively similar behavior. }
\label{Fig:b}
\end{figure}

In the previous section we introduced path integral approach to calculate power spectrum moments $M_k$ summing over the Dyck paths weighted by products of $b_n$. This approach immediately shows that if $b_n$ is a smooth function of $n$, at least for large $n$, then $M_k$ smoothly depends on $k$ for large $k$. Conversely, Lanczos coefficients $b_n$ can be calculated from from $M_k$ using the following relation 
\bea
\prod_{i=1}^n b_i^2=b_1^2\dots b_n^2 ={{\rm det}\, {\mathcal M}_{n+1}\over {\rm det}\,{\mathcal M}_{n}},
\eea
where ${\mathcal M}_{n}$ is a $n\times n$ Hankel matrix 
\bea
({\mathcal M}_{n})_{ij}=\left\{ 
\begin{array}{rc}
M_{i+j-2},& i+j\, {\rm mod}\, 2=0,\\
0, & i+j\, {\rm mod}\, 2=1.
\end{array} 
\right.
\eea
This expression allows calculating individual $b_n$ as a ratio of determinants, but it does not guarantee that $b_n$ will smoothly depend on  the index, even if $M_k$ do. 
To illustrate that smoothness of $M_k$ does not imply smoothness of $b_n$ we consider a mock autocorrelation function 
\bea
\label{mock}
C(-i\beta)={1\over 2}\left(e^{m (e^{\beta}-1)}+e^{m (e^{-\beta}-1)}\right), \label{m}
\eea
inspired by \eqref{answer1}. In this case $M_{2k}=B_{2k}(m)$ and Lanczos coefficients can be calculated numerically. They exhibit a peculiar behavior: initially $b_n$ seems to be a smooth function of $n$, but starting at some critical  $m$-dependent value behavior of $b_n$ for even and odd $n$ becomes drastically different. For even $n$,  $b^2_n\propto n^2$, while for odd $n$, $b_n^2\propto n$. 
This is shown in Fig.~\ref{Fig:b}. It should be noted that while mock correlation function \eqref{mock} exhibits expected behavior along the imaginary axis $t=-i\beta$, its behavior along real axis is periodic and hence unphysical. Thus, it remains to be seen if for lattice models with local interactions $b_n$ is always asymptotically smoothly depend of $n$, or the behavior can be more complicated.

\end{document}